\documentclass[fleqn,12pt,twoside]{article}
\usepackage{espcrc1}
\usepackage{amssymb, amsbsy}
\newif\ifpdf
        \ifx\pdfoutput\undefined
        \pdffalse 
        \else
        \pdfoutput=1 
        \pdftrue
        \fi
\ifpdf
        \usepackage[pdftex]{graphicx}
        \DeclareGraphicsExtensions{.pdf, .jpg}
\else
        \usepackage[dvips]{graphicx}
                \DeclareGraphicsExtensions{.eps, .jpg}
\fi

\def\vec#1{\boldsymbol{#1}}
\def\jipsi{\ensuremath\mathrm{J}/\psi}
\newcommand{\D}{\ensuremath\mathrm{D}}
\newcommand{\Db}{\ensuremath\overline{\mathrm{D}}{}}
\newcommand{\B}{\ensuremath{\mathrm{B}}}
\newcommand{\K}{\ensuremath{\mathrm{K}}}
\newcommand{\Kb}{\ensuremath\overline{\mathrm{K}}{}}

\newcommand{\Xmass}{\ensuremath{\mathrm{X(3872)}}}
\newcommand{\e}{\ensuremath{\mathrm{e}}}
\newcommand{\p}{\ensuremath{\mathrm{p}}}
\def\tso{\ensuremath{{}^3\mathrm{S}_1}}
\def\tdo{\ensuremath{{}^3\mathrm{D}_1}}
\def\tdt{\ensuremath{{}^3\mathrm{D}_2}}

\def\spo{\ensuremath{{}^1\mathrm{P}_1}}
\def\tpz{\ensuremath{{}^3\mathrm{P}_0}}
\def\tpo{\ensuremath{{}^3\mathrm{P}_1}}
\def\tpt{\ensuremath{{}^3\mathrm{P}_2}}

\newcommand{\AmS}{{\protect\the\textfont2
  A\kern-.1667em\lower.5ex\hbox{M}\kern-.125emS}}
\title{Recent issues in hadron spectroscopy%
\thanks{Based on an invited talk at the session on ``Subnucleon degrees of freedom'' at the International Nuclear Physics Conference, G\"oteborg, Sweden, June 2004}
}
\author{Jean-Marc~Richard\address[LPSC]{Laboratoire de Physique Subatomique et Cosmologie,\\ 
        Universit\'e Joseph Fourier-- CNRS-IN2P3,\\
        53, avenue des Martyrs, 38026 Grenoble cedex, France}%
}
\begin{document}
\maketitle
\begin{abstract}
A brief survey is presented of recently discovered hadrons, some of them  presumably demonstrating a new kind of internal structure. This includes: spin-singlet quarkonium, mesons with unexpected mass or width, baryons with two heavy quarks, and pentaquark candidates. Flavour configurations with a combination of light and heavy quarks appear as particularly promising.
\end{abstract}
\nocite*
\section{Introduction}
During recent months, several new hadrons have been discovered or suggested in a variety of experiments, triggering renewed phenomenological and theoretical studies.
The limits of the simple quark model have been reached and nowadays, understanding hadron spectroscopy requires  chiral dynamics, coupling to decay channels,  higher Fock configurations and, perhaps,  explicit gluon degrees of freedom.
Accordingly, potential models are not anymore tuned to reproduce the latest results accurately. They instead serve as a  guide to point out which states deviate from the main stream and call for a special treatment with new dynamical ingredients.

This contribution is devoted to a non-exhaustive survey of new hadrons, including: the radially excited charmonium $\eta'_c$, the $\D_s^*$ excitations of $\D_s$, the $\Xmass$, and the various  pentaquark candidates for pentaquark.

Newly discovered hadrons have motivated new models, or a renewed interest in existing models, that might address  some old problems, such as the abundance of light scalar mesons or the low mass of radially excited nucleons.

A major problem arises, indeed,  when counting scalar mesons below 2 GeV, as $u\bar{u}$, $d\bar{d}$ and $s\bar{s}$ states in a $\tpz$ state do not suffice to account for all observed states, even if one includes radial excitations. Multiquark states, meson--meson molecules, hybrids or glueballs are needed, with some amount of mixing with $q\bar{q}$ states. An argument in favour of multiquark was proposed in the framework of the MIT bag model \cite{JaffeScalar}, in which it does not cost more to create an additional  $s$-wave  $q\bar{q}$ pair than to promote an existing $q\bar{q}$ pair to a relative $\ell=1$ angular momentum. Clearly, a $(qq\bar{q}\bar{q})$ state can also be viewed as a compound meson--meson state, especially in the weak-binding regime. Some scalars have been described as $\K\Kb$ molecules. It was explained, e.g., by Isgur \cite{Geiger:1992va}  that various contributions to $q\bar{q}\leftrightarrow qq\bar{q}\bar{q}$ tend to cancel out, except in the scalar sector. So we have a possible understanding why scalar mesons proliferate more  than, e.g., vector ones.

The first sector where the constituent  quark model was taken seriously is that of baryon resonances, for which abundant data existed already in the early 60's. The paper by Greenberg and the lectures by Dalitz \cite{Greenberg:1964pe} on the harmonic-oscillator model has been followed by tens of more refined studies, including the much-celebrated work by Isgur and Karl \cite{IK}, who saw in the pattern of splittings strong evidence for chromomagnetic interaction at short distances.
Still, problems remain. Experimentally, some states are found with a mass lower than expected in naive potential models. Theoretically, models have been improved with relativistic kinematics, and coupling to decay channels, inducing also mass shifts. Spin--flavour dynamics has been proposed, as an alternative to chromomagnetism (spin--colour). Lattice QCD suggests that the hierarchy of levels, as it is observed, is an effect of the very light mass of $u$ and $d$ quarks. For higher (unphysical) values of the light-quark mass, one would get the quarkonium type of ordering, with the orbital excitation lower than the radial one. For years, the three-body dynamics of the harmonic-oscillator or more refined potential models has been challenged by diquark models, where a baryon is a bound state of a quark and a diquark.  These latter models do not predict as many baryon resonances as the ordinary quark model. For instance, the  $[20,1^+]$ multiplet where both degrees of freedom $\vec{\rho}\propto\vec{r}_2-\vec{r}_1$ and $\vec{\lambda}\propto2\vec{r}_3-\vec{r}_1-\vec{r}_2$ are excited, exists in  the conventional model: these states are weakly coupled to the traditional doorway channels $\pi+N$ or $\gamma+N$, but should show up somehow in high-statistics experiments. On the other hand, this state does not exist in the simple diquark model with s-wave diquarks.  The diquark model has several other appealing features but  lacks  a strict derivation in the framework of QCD.
\section{Charmonium}
A recent review of new results on heavy quarkonia is given in Ref.~\cite{Stoeck:2004bs} and in the ``Yellow Report'', to appear shortly, of the \emph{Quarkonium Working Group} \cite{qwg}. Of interest, for instance, is the discovery of a \tdt\ state in the Upsilon family. The non-observation in $\e^+\e^-$ of any \tdo\ of $(b\bar{b})$ confirms that if the $\psi''$ state of  the charmonium family is seen formed in $\e^+\e^-$, this is due to its $\tso$ mixing induced by tensor forces, suppressed by 
$(m_c/m_b)^2$ for Upsilon, and to the  particular coupling induced by the neighbouring $\D\Db$, a situation without analogue for the 2S level of $(b\bar{b})$. The influence of $\tso\leftrightarrow\tdo$ mixing on the  decay properties is discussed, e.g., in \cite{Rosner:2004mi}.

Orbital excitations of mesons have been known for many years. The nice alignment of the square mass as a function of the angular momentum motivated the phenomenology of Regge trajectories, which was a source of inspiration for the development of string theory.

The mesons named ``radial excitations''  in the quark-model language, correspond to the ``daughter'' trajectories in the Regge terminology, which are less easily produced. The first clean evidence of the radial degree of freedom shows up in the charmonium spectrum, with a narrow $\psi'$ with the same quantum numbers as $\jipsi$, and  similar decay properties, except for the cascade decay modes $\jipsi + X$.

The $\psi(4040)$ has an interesting history. It was found to have preferential decay into $\D\Db^*$ (an implicit ${}+ {\rm c.c.}$ is implied here and in similar circumstances). This suggested a molecular description of this state \cite{Voloshin:1976ap}. In fact, the groups at Orsay  and  Cornell  understood that the unorthodox pattern of  branching ratios to $\D\Db$, $\D\Db^*$ and $\D^*\Db^*$ is due to the radial structure of $\psi(4040)$, as a mere $c\bar{c}$ state \cite{LeYaouanc:1977gm}. The spatial wave function has nodes. In momentum space, there are also oscillations. Hence, if a decay calls for a momentum whose probability is low, it is suppressed.

Once the situation clarified for the $\psi(4040)$, one gets a reasonable sequence of radial excitations for the spin-triplet charmonium states: $\jipsi$, $\psi'$,  $\psi(4040)$, etc., and a similar and longer sequence in the Upsilon family. The spacing pattern is not rigorously monotonic, as it should be in a monotonic confining potential, but this is due to the opening of thresholds or $\tso-\tdo$ mixing, etc., which induce shifts of the energy levels.

The search for spin-singlet states was anticipated to be difficult. Radiative decay (M1 transition) from the same multiplet (e.g., $\Upsilon\to\eta_b+\gamma$) are suppressed by the small energy release, and those from higher multiplets (e.g., $\psi'\to\eta_c+\gamma$) are hindered by the orthogonality of the wave functions. The search in $\rm{\bar p}p$ collisions requires paradoxically the prior knowledge of the precise mass, and this results in tedious and speculative scans. It remains that the R704 experiment at CERN, the E760 and E835 at Fermilab have made major contributions to charm physics, as the future Panda experiment certainly will do. It is a nice surprise that the $\eta_c$ is seen formed in $\rm{\bar p}p$ collisions, but the coupling is perhaps less favourable for the $\eta_c'$, and the search in the last $\rm{\bar p}p$ runs was concentrated on the lower part of the plausible mass interval. 

The embarrassing absence of reliable $\eta'_c$ is probably over now, with the observation by the Belle collaboration of $\eta'_c$ in two different measurements \cite{BelleEta2S}. The first one is double-charm production in $\e^+\e^-$ collisions, that is to say, $\jipsi + X$. The Zweig rule strongly suggests that $c\bar{c}$ recoils against $c\bar{c}$. There is, however, some debate on the cross-section for these double-charm events.  The second Belle measurement deals with \B\ decay. The final state $\K\K\K\pi$ exhibits a peak in the $\K\K\pi$ mass spectrum. The $\eta'_c$ has also been seen in other experiments.

The $\eta'_c$ is closer to $\psi'$ than expected in simple charmonium models. This is probably due to a coupling to the open-charm thresholds. The $\eta_c'$ does not link to $\D\Db$, due to its pseudoscalar nature, but the $\psi'$ does, and is pushed down. This was noted in Ref.~\cite{Martin:1982nw}, on the basis of the Cornell model.

The $\spo$ state of charmonium is searched for desperately. It was suspected in the R704 experiment at CERN ISR, and seen for a while in the Fermilab $\rm{\bar p}p$ experiment, but not confirmed by the last runs. The mass was just where it was most naively expected, at the centre of gravity of \tpz, \tpo\ and \tpt\ triplet states. The $\spo$, indeed, gives the opportunity to test the short-range character of the spin--spin component of the quark--antiquark interaction.
\section{New mesons}
The most exciting state in the charmonium-mass range is the $\Xmass$. It is seen in several experiments, as a clear signal above  the background, and hence its existence can be considered as very safely established. See, e.g., \cite{Aubert:2004ns} and references there. Note it is not seen in two-photon production at CLEO, this constraining the possible quantum numbers.

Though experts have learned to be careful from the lesson of the $\psi(4040)$, see above, a molecular interpretation of this state is very tempting:\\[-20pt]
\begin{itemize}\itemsep-3pt
\item
It lies almost exactly at the $\D^0\Db^{*0}$ threshold.
\item
None of the charmonium assignment (D-state, radially excited P-state, etc.) survives scrutiny.
\item
It was predicted with this structure by T\"ornqvist on the safe ground (especially for this audience of nuclear physicists) of pion-exchange interaction between two hadrons \cite{Tornqvist:2003na}.
\end{itemize}\vskip -4pt
If this picture is true, we have here the analogue of a slightly unbound nucleus, with nucleons replaced by charmed mesons. Other contributors to the ``nuclear physics'' of heavy-flavoured hadrons include
Ericson and Karl, Manohar and Wise, Braaten, and more recently Julia-Diaz and Riska \cite{Ericson:1993wy}. In most of these calculations, one finds the effective hadron--hadron potential at the edge between binding and non-binding. This is an opportunity to remind hadron physicists about the possibility of \emph{Borromean binding}~\cite{Richard:2003nn}: if two particles do not bind, the same pairwise interaction can well stabilise a three-body system. For instance, in nuclear physics, neither the $\alpha-\mathrm{n}$ (${}^5$He) nor the n--n system have any bound state, but the three-body combination ($\alpha-\mathrm{n}-\mathrm{n}$) gives the stable ${}^6$He. 

One may wonder why the Yukawa potential, barely strong enough to bind two nucleons, can give enough attraction for charmed particles, which contain less light quarks. The potential is, indeed, weaker, but it is experienced by heavier particles, which take more benefit of the attractive part. The key quantity for binding  is the product of the reduced mass by  the strength of the interaction. 

Whilst a consensus seems to emerge for the $\Xmass$, the $\D_{s,J}$ states remain more mysterious. Let me summarise briefly the story, first on the experimental side. 
States with $(c\bar{s})$ flavour content were found at the Babar experiment of SLAC, and confirmed as CLEO, Fermilab, etc.~\cite{Bondioli:2004te}.
The masses are $2317$ and $2458\;\mathrm{MeV}$, to be compared with 1968 for the pseudoscalar and $2112\;\mathrm{MeV}$ for the vector ground state of $(c\bar{s})$. This is rather low for orbital excitations, from our present understanding of spin-orbit forces. 
But the main problem is with the widths, which are very small.
There are schematiically two schools.\\[-20pt]
\begin{enumerate}\itemsep -3pt
\item
These new states are  understood as the chiral partners of the ground-state multiplet.  Some authors insist of that this is not an ad-hoc explanation just for these two states, but a recurrent phenomenon for light quarks surrounding an heavy core \cite{Bardeen:2003kt}.
\item
Barnes et al.\ \cite{Barnes:2003dj}, and several other authors proposed a four-quark interpretation,  $(cq\bar{s}\bar{q})$, where $q$ denotes a light quark, $u$ or $d$, with however, a possible breaking of isospin symmetry.
\end{enumerate}\vskip -4pt
 It would be necessary to investigate in detail the charm--strange sector to see whether the new $\D_{s,J}$ states are just the usual excitations, shifted in mass and made narrower than expected, or supernumerary states.

Shortly before this conference, a preprint was released  by the Selex collaboration, announcing a new state, $D_{s,J}(2632)$, again rather narrow, and decaying more often into $\D_s\eta$ than into $\D\mathrm{K}$. We had  probably at G{\"o}teborg the first official presentation of the $\D_s(2632)$, in the previous talk, by Jurgen Engelfried \cite{Engelfried}.
In between this conference and the writing of the Proceedings, there have been several papers on this
puzzling state, in which some colleagues see the baryonium striking again.
\section{Double-charm baryons}
The experimental situation on this sector has been reviewed by the previous speaker~\cite{Engelfried}. It is extremely interesting that the lowest $\Xi^+_{cc}(ccd)$ state is now seen in two different weak-decay modes. Problems of course remain: Selex is the only experiment having seen these baryons, yet; the isospin splitting between this $\Xi^+_{cc}$ and the lowest $\Xi^{++}_{cc}$ candidate is larger than expected; the puzzling excitations about 60 MeV or so above the ground state need confirmation with higher statistics.

Anyhow, the results of the Selex collaboration have stimulated further studies on hadrons with two heavy quarks. The $QQq$ baryons are perhaps the most interesting of ordinary hadrons, as they combine in a single object two extreme regimes: the slow motion of two heavy quarks in an effective potential generated by light degrees of freedom, as in charmonium or Upsilon systems; the ultra-relativistic motion of a light quark around a heavy colour source, as in $\D$ and $\B$ mesons.

In $(QQq)$, the $QQ$ separation is certainly smaller than the $Qq$ one. This suggests some simplified treatment of the dynamics. In several papers, a $QQ$ ``diquark'' is first estimated which is bound to a light quark in a second step. One should keep in mind that the first excitation will show up inside the heavy diquark, and also that the $QQ$ interaction is an effective one, influenced that the light quark. For instance, in the harmonic-oscillator model, the bare $QQ$ potential gains $50\%$ increase by the presence of the light quark. For the diquark approach, see, e.g.~\cite{Ebert:2002ig}.

Another approach is the Born--Oppenheimer approximation \cite{Fleck:1989mb}, similar to the treatment of H$_2^+$ in atomic physics. For a given $QQ$ separation, one solves the two-centre problem for the light quark, and this energy, supplemented by the direct $QQ$ interaction, generated the effective $QQ$ interaction. This method was tested with simple potential models. It could be applied with a better treatment of the light quark dynamics, for instance using lattice simulations.

As pointed out \cite{Bardeen:2003kt}, one expects chiral partners of $(QQq)$ baryons, as we seemingly observe chiral partners of $(Q\bar{q})$ mesons. However, an universal spacing of about $300\;$MeV is empirically observed between any hadron and its chiral partner. Hence the small spacing suggested by the Selex results is difficult to understand.
\section{Double-charm exotics}
The mechanism leading to $\D\Db^*$ molecules also applies to charm exotics ($C=2$): in some spin--isospin configurations, the long-range potential mediated by pion exchange between two D$^{(*)}$  is attractive.

At the quark level,  the $(QQ\bar{q}\bar{q})$ configuration benefits from a favourable binding as compared to its threshold $(Q\bar{q})+(Q\bar{q})$ \cite{Richard:1999qh}.
Again, heavier particles, namely the $QQ$ pair, take more advantage of an attractive  interaction. This effect is well known in atomic physics: the same Coulomb potential produces a marginally bound Ps$_2$ molecule $(\e^+,\e^+\,e^-,\e^-)$, and a more deeply bound H$_2$ molecule $(\p,\p,\e^-,\e^-)$.

When taken independently, the long-range $\D\Db^*$ interaction or the $(cc\bar{q}\bar{q})$ four-quark dynamics are not strong enough to safely produce a bound state. However, 
it is presumably possible to find a configuration where both Yukawa forces and short-range quark forces cooperate. Hence exotic  mesons with charm $C=2$ await discovery. 
\section{Light pentaquark}
A baryon with charge $Q=+1$ and strangeness $S=+1$, i.e., minimal quark content $(\bar{s}uudd)$ has been seen at the Spring8 facility in Japan, and in several other experiments.  On the other hand, this state is not seen in a number of high-statistics experiments. For a review, see \cite{Rossi:2004rb}. The least one can say, is that the situation is rather uncertain and requires further scans. One may wonder why collaborations having data on tape for years never had a look before at exotic hadrons, till this became a fashionable topic. These are high-quality  experiments, deserving a full use of the recorded events.

A fraction of the NA49 collaboration at CERN published data with a peak in a configuration with strangeness $S=-2$ and charge $Q=-2$, corresponding to  $(\bar{u} ssdd)$. This state, however, resists the scrutiny of other members of this collaboration, and the analysis of other data sets, and thus is controversial.

It is also proposed that some baryon resonances with ordinary quantum numbers contain a large fraction of pentaquark configurations, to explain their intriguing properties. This concerns, e.g., the Roper resonances and the $\Lambda(1405)$.

On  the theory side, the $S=+1$ pentaquark at about $1.5\;$GeV was predicted by Diakonov at al., following other pioneering works on the chiral soliton dynamics and in particular the Skymion model \cite{Diakonov:2004ie}. The remarkable feature of this approach is the existence of an antidecuplet ($\overline{10}$) of baryons on almost the same footing as the familiar octet (nucleon, $\Lambda$, \dots) and decuplet ($\Delta$, \dots, $\Omega^-$).

In most other approaches, one gets more easily a negative parity. Since in the Dirac theory, an antiparticle has an intrinsic parity opposite to that of the associated particle, a $(\bar{q}q^4)$ with relative  s-wave among the constituents has a parity opposite to that of a $(qqq)$ ground-state baryon.

An exception is the model by Jaffe and Wilczek (see \cite{Jennings:2003wz} and refs.~there), involving two colour $\bar{3}$, spin 0, $[ud]$ diquarks, which have a relative orbital momentum $\ell=1$ to obey Bose statistics, as their colour coupling $\bar3\times\bar3\to3$ is already antisymmetric. Lattice and sum-rule calculations assuming the same diquark structure also get a positive-parity pentaquark, not surprisingly. Most other lattice calculations predict a negative parity for this configuration, or no bound state or resonance at all \cite{LatticePenta}.

Another exception is proposed by Stancu and Riska, who used the spin--flavour interaction of Glozman et al., which has some phenomenological success in describing the spectrum of baryon excitations \cite{Stancu:2003if}. Due to the state-dependence of the interaction, the lowest p-wave tends paradoxically to become the ground-state. This is somewhat similar to the pattern observed years ago for pions bound to nuclei. 

In most models explaining the existence of  a light pentaquark with strangeness $S=+1$, a heavy $(\overline{Q}qqqq)$ version exists, tentatively more stable against dissociation. Heavy pentaquarks have already a long history. A first candidate was proposed in 1987 independently in a paper by the Grenoble group, where the word ``pentaquark'' was seemingly used for the first time, and another one by Lipkin \cite{Gignoux:1987cn}. It consists of $(\overline{Q}qqqq)$, with the light--strange sector $q^4$ form a flavour triplet of SU(3). The parity is negative in the original model, very much inspired from Jaffe's H$(uuddss)$, the binding being due to attractive coherences in the chromomagnetic interaction.  
\section{Other multiquarks}
The mechanisms astutely designed to explain the $X(3862)$, the $\D_s^*$, and the tentative pentaquark offer the fascinating risk of predicting other multiquarks, and this is already stimulating several experimental searches. It thus seems timely to review the state of art in this field.

Multiquark spectroscopy is almost as old as meson and baryon spectroscopy in the quark model. It is regularly abandoned and reactivated, except in the context of constituent quark models, where it is used a playground for elaborated few-body calculations.
\subsection{Potential models}
It is hardly justified to use potential models for computing hadron masses and properties, except perhaps for the lowest levels of systems containing solely heavy quarks, for which an adiabatic limit can be approached from basic QCD~\cite{qwg}.

An additional difficulty arises when one tries to extrapolate the empirical interquark potential to systems containing more quarks. From $N=2$ constituents (quark--antiquark mesons) to $N=3$ (three-quark baryons), the so-called ``1/2 rule'', i.e., 
\begin{equation}\label{eq:half}
V_{QQQ}={1\over 2} \sum_{i<j}V_{Q\overline{Q}} (r_{ij})~,
\end{equation}
is similar to the rule $V_{I=0}=-3 V_{I=1}$ for a component of the nucleon--nucleon potential in isospin $I$ arising from isovector exchange. 
Equation (\ref{eq:half}) would correspond to pairwise forces with pure colour-octet exchange. It can, however, be justified as an approximation to a ``$Y$-shape'' string linking the three quarks.
The generalisation to more complicated multiquark systems
\begin{equation}\label{eq:lambda}
V=-{16\over3} \sum_{i<j} \tilde\lambda^{c}_i.\tilde\lambda^{c}_j\,v(r_{ij})~, 
\end{equation}
where $v$ is an empirical potential, is even more adventurous. Still, it provides a toy model, where certain effects can be tested.  Among the lessons, one can underline:
\begin{enumerate}\itemsep-2pt
\item Baryons are heavier than mesons when the mass is counted per quark, i.e., $(qqq)/3\ge (q\bar{q})/2$. This explains, e.g., why a baryon--antibaryon system can annihilate by mere rearrangement
$q^3+\bar{q}{}^3\to 3 (q\bar{q})$.
\item The tendency seemingly persists, i.e., $M_N/N\nearrow$ for a system of $N$ quarks and antiquarks. In particular, $(q^2\bar{q}^2)\ge 2 (q\bar{q})$.
\end{enumerate}
In other words, there is \emph{no proliferation} of multiquarks in the simple model (\ref{eq:lambda}), and  binding (or nearby resonance) requires special additional terms in the Hamiltonian. Among the routes that have been explored, one may mention, besides the already mentioned long-range forces and quark-mass asymmetries:
\begin{itemize}\itemsep-2pt
\item%
Chromomagnetism. This is the main ingredient of early models of the $\mathrm{H}(uuddss)$ or heavy pentaquarks $\mathrm{P}(\overline{Q}qqqq)$. Schematically,\\[-10pt]
\begin{equation}\label{eq:spin-colour}
V_{SS}=-\sum_{i<j}C\delta^{(3)}(r_{ij}) \vec{\sigma}_i . \vec{\sigma_j}\,\tilde\lambda^{c}_i.\tilde\lambda^{c}_j~,
\end{equation}
\vskip -8pt
If the strength $C$ and short-range correlation $\langle \delta^{(3)}(r_{ij})\rangle$ are  the same among all light quarks (or antiquarks) $u$, $d$ and $s$, and is assumed to be as strong as for ordinary hadrons (where its value is directly linked to hyperfine spacings), then a binding as large as $B=-150\;$MeV is predicted for the $H$, and for the $P$ in the limit where $m(Q)\to\infty$. However, SU(3)$_F$ breaking, and any serious treatment of the 5- or 6-body dynamics leaves the $\mathrm{H}$ or $\mathrm{P}$ unbound. In particular $\langle \delta^{(3)}(r_{ij})\rangle$ is much smaller in a dilute multiquark than in a compact baryon.
\item%
Spin--flavour. The functional dependence is similar to Eq.~(\ref{eq:spin-colour}), but the colour operator $\lambda^c$ is replaced by its flavour analogue. There is no proliferation of multiquarks in this model, mainly due to the weakening of the short-range correlation in composite systems, as for the chromomagnetic model. An interesting difference with respect to (\ref{eq:spin-colour}) is the lowering of negative-parity pentaquark states.
\end{itemize}

One should add that even for simple non-relativistic calculations with pairwise forces, \emph{ a minimal few-body expertise} is required. In particular, one should control the dissociation threshold, and interpret with care any ``energy'' lying above the threshold. An expectation value $\langle \Psi\vert H\vert \Psi\rangle=M_1+M_2+100\;$MeV, for instance, does not mean one has a resonance 100 MeV above the threshold $M_1+M_2$. It simply means that the trial wave function is too crude to reproduce a threshold made of well-separated hadrons 1 and 2. The same warning holds for more refined multiquark calculations such as relativistic models, lattice simulation, sum rules, etc.

Note that if a system (in a model) is at the edge of binding, its wave function extends very far outside the interaction region.  Truncating the tail is equivalent to adding an unphysical external hard core, which spoils the binding.
\subsection{Diquark compounds}
Once diquarks are introduced, with an assumed mass that is low enough to explain the low mass of $\theta(1540)$ as $\bar{s}[ud][ud]$ with $\ell=1$, one may wonder whether diquarks will manifest themselves elsewhere.

For instance, a tri-diquark $[ud]^3$ with two units of orbital momentum will produce a $J^P=1^1$ ``demon-deuteron'' which is likely below the normal deuteron~\cite{Fredriksson:1981mh}. If diquark clustering is not restricted to $[ud]$, then a $[ud][ds][su]$ state, in S-wave, would resuscitate a bound $\mathrm{H}$ below the $\Lambda\Lambda$ threshold, without the need for coherences in the spin--colour terms. See, also, \cite{Zhu:2004za}.

A triquark $[\bar{s}ud]$ has even been proposed \cite{Karliner:2003dt} to supplement the $[ud]$ diquark, making the putative $\theta(1540)$ a quasi two-body system $[\bar{s}ud][ud]$. If the same clustering is also at work in larger systems, one may get surprises, such as $[\bar{s}ud]^3$, which could manifest itself as a bound state of the $\overline{\Omega^-}$-deuteron system.

One should thus be careful with diquarks, as some of their most steadfast promoters are \cite{Lichtenberg:2004tb}, and refrain from opening a new Pandora's box of many multiquarks, years  after the baryonium episode. Diquarks can either be very enlightening or very misleading. For instance, more than half a century ago, it was noticed that the slope of the Regge trajectories is the same for mesons and baryons. Then it was pointed out  that if one \emph{assumes} a quark--diquark structure of baryons, the property becomes natural, as for both mesons and baryons, the same string is excited, that links a colour 3 to a colour $\bar{3}$ source. But it took years to \emph{show} that this clustering spontaneously arises in a large class of models: when one implements an angular momentum $\ell$ between three quarks bound by a linear confinement (or its $Y$-shape improved version), the minimal energy, for large $\ell$, is obtained when a single quark rotates around the two others remaining in a relative $s$-wave~\cite{Martin:1985hw}.  On the other hand, the failure of the simple quark-model of baryonium comes that it was \emph{postulated} that orbital excited $(qq\bar{q}\bar{q})$ could consist of a diquark $[qq]$ well separated from an antidiquark. The game was even more tantalising with colour-sextet diquarks. However, no dynamical model ever explained why this was the appropriate structure.
\section{Conclusions}
The last months were very exciting in the forehead of hadron spectroscopy. The $\D_{s,J}$ resonances, the heavy baryon, the $\Xmass$, and the pentaquark candidates stimulated interesting studies on confinement dynamics.

The discovery of the $\eta'_c$ shows that new means of investigation can solve old problems, and it is hoped that missing baryons and quarkonia will also be found with an appropriate production mechanisms.  

For many years, hadrons with multiquark structure, or constituent glue, or revealing the power of chiral symmetry have been searched for. Now, they are emerging perhaps too suddenly, and we would be rather embarrassed if all recent candidates would survive careful experimental scrutiny. We have to wait for the current next wave of experiments and analyses, especially concerning the controversial pentaquark.

Meanwhile, theorists should also refine and improve their tools. The history of poly-electrons  is in this respect rather instructive. Around the year 1945, Wheeler proposed several new states, in particular the $(e^+,e^+,e^-,e^-)$, as being stable if internal annihilation is neglected.  In 1946, Ore published an article where he concluded that stability is very unlikely, on the basis of a seemingly-solid variational calculation borrowed from a nuclear-physics picture of the $\alpha$-particle as a four-nucleon system. 
Hylleraas, however, suspected that the trial wave function was not suited for long-range forces. Today, most of us, in similar circumstances, would rush to their computer and post a criticism of the web. These gentlemen, instead, combined their efforts, and in 1947, published a very elegant and rigorous proof of the stability \cite{Whe46},
\subsection*{Acknowledgments}
It is a pleasure to thanks Bj\"orn Jonson and his colleagues for this beautiful conference, D.O.~Riska for the lively session on hadrons, several colleagues for informative discussions  and, in particular, E.~Leader for comments on the manuscript.

\end{document}